# Associative Arrays: Unified Mathematics for Spreadsheets, Databases, Matrices, and Graphs

Jeremy Kepner[1,2,3,4], Julian Chaidez[1,4], Vijay Gadepally[2,3,4], Hayden Jansen[1,4]
[1]MIT Mathematics Department, Cambridge, Massachusetts
[2]MIT Computer Science & AI Laboratory, Cambridge, Massachusetts
[3]MIT Lincoln Laboratory, Lexington, Massachusetts
[4]MIT BeaverWorks Center, Cambridge, Massachusetts

*Abstract*—Data processing systems impose multiple views on data as it is processed by the system. These views include spreadsheets, databases, matrices, and graphs. The common theme amongst these views is the need to store and operate on data as whole sets instead of as individual data elements. This work describes a common mathematical representation of these data sets (associative arrays) that applies across a wide range of applications and technologies. Associative arrays unify and simplify these different approaches for representing and manipulating data into common two-dimensional view of data. Specifically, associative arrays (1) reduce the effort required to pass data between steps in a data processing system, (2) allow steps to be interchanged with full confidence that the results will be unchanged, and (3) make it possible to recognize when steps can be simplified or eliminated. Most database system naturally support associative arrays via their tabular interfaces. The D4M implementation of associative arrays uses this feature to provide a common interface across SQL, NoSQL, and NewSQL databases.

Keywords-Insider; Big Data; Associative Arrays; Spreadsheets; Database; Matrices; Graphs; Abstract Algebra

## I. Introduction

As data moves through a processing system the data are viewed from different perspectives by different parts of the system (see Figure 1). Data often are first parsed into a tabular spreadsheet form (e.g., .csv or .tsv files), then ingested into database tables, analyzed with matrix mathematics, and presented as graphs of relationships. A large fraction of the effort of developing and maintaining a data processing system goes into sustaining these different perspectives. It is desirable to minimize the differences between these perspectives. Fortunately, spreadsheets, databases, matrices, and graphs all use two-dimensional data structures in which each data element can be specified with a triple denoted by a row, column, and value. Using this common reference point, many technologies have been developed to bridge the gaps between these different perspectives. Array programming languages (e.g., Matlab, R, and Python) have been the *de facto* standard for manipulating matrices (both dense [Moler 1980, Moler 2008] and sparse [Gilbert, Moler & Schreiber 1992]) since the 1990s. These languages have had direct support for spreadsheet manipulation for nearly as long. An even stronger connection exists between spreadsheets and relational databases. A prime example is the SAP enterprise resource planning package (www.sap.com), which is the dominant software used for accounting and payroll management throughout the world. SAP relies on seamless integration between SQL databases and spreadsheets. More recently, spreadsheets have incorporated adjacency matrices to manipulate and visualize graphs by using their built in scatter plotting capabilities [Smith et al 2009]. Perhaps the largest recent development has been the introduction of key-value store databases [Wall, Cordova & Rinaldi 2013], which are specifically designed to store massive sparse tables and are ideal for storing graphs. Array store databases [Balazinska et al 2009] have taken sparse tables a step further by also including first-class support of matrix operations on that data. The deep connection between graphs and sparse matrices [Kepner & Gilbert 2011] has been recognized to such an extent that it has led to the development of the GraphBLAS standard for bringing these fields together [Mattson et al 2013, Mattson 2014, Gilbert 2014, Kepner & Gadepally 2014, Buluc et al 2014].

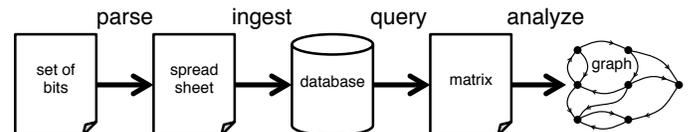

Figure 1. The standard steps in a data processing system often require different perspectives on the data. Associative arrays enable a common mathematical perspective to be used across all the steps.

The D4M software system (d4m.mit.edu) [Kepner 2011p, Kepner et al 2012] is the first practical implementation of associative arrays that successfully bridges spreadsheets, databases, matrices, and graphs. Using associative arrays, D4M users are able to implement high performance complex algorithms with significantly less effort. In D4M, a user can read data from a spreadsheet, load the data into a variety of databases, correlate rows and columns with matrix operations, and visualize connections using graph operations. These operations correspond to the steps necessary to build an end-to-end data processing system. Often, the majority of time spent in building a data processing system is in the interfaces between the various steps. These interfaces require a conversion from one mathematical perspective on the data to another. By using

This material is based upon work supported by the National Science Foundation under Grant No. DMS-1312831. Any opinions, findings, and conclusions or recommendations expressed in this material are those of the author(s) and do not necessarily reflect the views of the National Science Foundation.

a common mathematical abstraction across all steps, the construction time of a data processing system can be reduced.

Relational or SQL (Structured Query Language) databases [Codd 1970, Stonebraker et al 1976] have been the de facto interface to databases since the 1980s and are the bedrock of electronic transactions around the world. More recently, key-value stores (NoSQL databases) [Chang et al 2008] have been developed for representing large sparse tables to aid in the analysis of data for Internet search. As a result, the majority of the data on the Internet is now analyzed using key-value stores [DeCandia et al 2007, Lakshman & Malik 2010, George 2011]. In response to the same challenges, the relational database community has developed a new class of array store (NewSQL) databases [Stonebraker et al 2005, Kallman et al 2008, Lamb et al 2012, Stonebraker & Weisberg 2013] to provide the features of relational databases while also scaling to very large data sets.

The diversity of databases has created a need to interoperate between them. Associative arrays provide an abstraction that works with all of these classes of databases (SQL, NoSQL, and NewSQL) and can be bound to database tables, views, or queries. D4M has demonstrated this capability [Wu et al 2014]. One example where this is useful is in the field of medicine, where a SQL database might be used for patient records, a NoSQL database for analyzing the medical literature, and a NewSQL database for analyzing patient sensor data.

The success of D4M in building real data processing systems has been a prime motivation for formalizing the mathematics of associative arrays. By making associative arrays mathematically rigorous, it becomes possible to apply associative arrays in a wide range of programming environments (not just D4M).

## II. ASSOCIATIVE ARRAY INTUITION

Associative arrays derive much of their power from their ability to represent data intuitively in easily understandable tables. Consider the list of songs and the various features of those songs shown in Figure 2. The tabular arrangement of the data shown in Figure 2 is an associative array (denoted **A**). This arrangement is similar to those widely used in spreadsheets and databases. Figure 2 illustrates two properties of associative arrays that are different from other two-dimensional arrangements of data. First, each row label (or row key) and each column label (or column key) in **A** is unique, which allows rows and columns to be queried efficiently. Second, associative arrays contain no rows or columns that are entirely empty, which allows insertion, selection, and deletion of data to be performed by associative array addition, multiplication, and products. These properties are what makes **A** an associative array and allows **A** to be manipulated as a spreadsheet, database, matrix, or graph.

| **A** | Artist | Date | Duration | Genre |
|---|---|---|---|---|
| 053013ktnA1 | Bandayde | 2013-05-30 | 5:14 | Electronic |
| 053013ktnA2 | Kastle | 2013-05-30 | 3:07 | Electronic |
| 063012ktnA1 | Kitten | 2010-06-30 | 4:38 | Rock |
| 082812ktnA1 | Kitten | 2012-08-28 | 3:25 | Pop |

Figure 2. Tabular arrangement of a list of songs and the various features of those songs into an associative array **A**. The array **A** is an associative array because each row label (or row key) and each column label (or column key) in **A** is unique.

## III. MATHEMATICAL OPERATIONS

Addition, multiplication, and products are the most commonly used operations for transforming data and also the most well studied mathematically. The first step in understanding associative arrays is to define what adding or multiplying two associative arrays means. Addition and multiplication of associative arrays have properties that are different from arithmetic addition (e.g., $1 + 2 = 3$) and multiplication (e.g., $2 \times 3 = 6$). To prevent confusion with arithmetic addition and multiplication, $\oplus$ will be used to denote associative array addition and $\otimes$ will be use to denote associative array multiplication.

Given associative arrays **A**, **B**, and **C**, associative array addition is denoted

$$C = A \oplus B$$

Associative array addition is equivalent to database table insertion in the formula

$$T = T \oplus B$$

where **T** is an associative array that is *bound* to a database table, view, or query. Associative array element-wise multiplication is denoted

$$C = A \otimes B$$

Associative array element-wise multiplication is equivalent to database table selection in the formula

$$C = T \otimes B$$

where **C** has the elements in **T** corresponding to the non-zero (or non-empty) entries in **B**. Associative array (matrix) product combines addition and multiplication and is written

$$C = A \ B$$

The above product can also be denoted $\oplus.\otimes$ to highlight its special use of both addition and multiplication

$$C = A \ \oplus.\otimes \ B$$

or

$$C(i,j) = \oplus_k \ A(i,k) \otimes B(k,j)$$

The special case of using associative array products for row selection is often denoted by using parentheses

$$T(a,:) = A \ T$$

where **a** are the columns of a permutation array **A** (see section V) that correspond to the rows to be selected from **T**. Likewise, column selection can be denoted

$$T(:,b) = T \ B$$

where **b** are the rows of a permutation array **B** that correspond to the columns to be selected in **T**.

One of the most interesting properties of an associative array is how sub-arrays are handled. Sub-arrays are extracted using ranges or sets of row and column keys. The row keys and column keys of non-empty rows and columns are carried along into the sub-array. Associative arrays also allow the same sub-array selection to be performed via either element-wise or matrix products. The duality between array selection and array products allows this operation to be treated in the same manner as other algebraic operations.

Row and column keys are always carried with the associative array and the associative array does not hold empty rows or empty columns. Thus, inserting or assigning values to an associative array can also be carried out via addition.

Products of associative arrays are one of the most useful operations in a data processing system. Associative array products can be used to correlate one set of data with another set of data, transform the row or column labels from one naming scheme to another, and aggregate data into groups. Figure 3 shows how different musical genres can be correlated by artist using associative array matrix products.

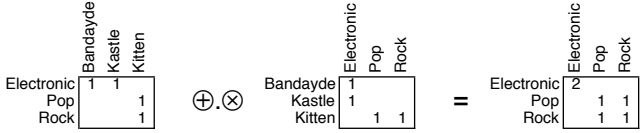

Figure 3. Correlation of different musical genres using associative array matrix product ⊕.⊗

Associative array addition, element-wise multiplication, and matrix product can be defined so that they are algebraically correct for spreadsheets, databases, matrices, and graphs. Algebraic rigor is what allows associative arrays to be an effective tool for manipulating data in a wide range of applications. The keys to defining these operations are basic concepts from abstract algebra that extend the ideas of addition, multiplication, and products to numbers and words.

## IV. FORMAL DEFINITIONS

Associative arrays are effective because it is possible to formally prove that for associative arrays of all shapes and sizes that addition, element-wise multiplication, and matrix products maintain their desirable algebraic properties [Kepner 2012, Kepner & Chaidez 2013, Kepner & Chaidez 2014, Kepner & Jansen 2015]. Perhaps the most important of these properties is coincidentally called *associativity,* which allows operations to be grouped

$$(A \oplus B) \oplus C = A \oplus (B \oplus C)$$
$$(A \otimes B) \otimes C = A \otimes (B \otimes C)$$
$$(A\ B)\ C = A\ (B\ C)$$

Associativity enables operations to be executed in any order and is extremely useful for data processing systems. The ability to swap steps or to change the order of steps in a data processing system can significantly simplify its construction. For example, if arrays of data are entering a system one row at a time and the first step in processing the data is to perform an operation across all columns and the second requires processing across all rows, this can make the system difficult to build. However, if the processing steps are associative, then the first and second steps can be swapped, making it much easier to build the system.

## V. SPECIAL ARRAYS AND GRAPHS

The internal structure of the associative array is important for a range of applications. In particular, the distribution of nonzero entries in an array is often used to represent relationships that can also be depicted as points (vertices) connected by lines (edges). These diagrams are called graphs. For example, one such set of relationships are those genres of music that are performed by particular musical artists. Figure 3 extracts these relationships from the data in Figure 2 and displays it as both an array and a graph.

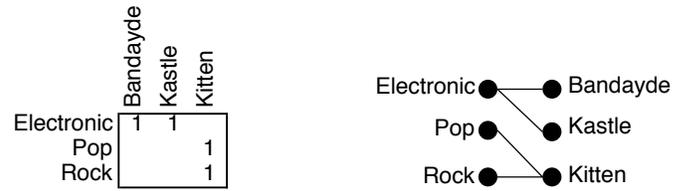

Figure 4. Relationships between genres of music and musical artists taken from the data in Figure 2. The array on the left shows how many songs are performed for each genre and each artist. The graph on the right shows the same information in visual form.

Certain special patterns of relationships appear frequently and are of sufficient interest to be given names. Modifying Figure 3 by removing some of the relationships (see Figure 4) produces a special array whereby each row corresponds to exactly one column. Likewise, the graph of these relationships shows the same pattern, and each genre vertex is connected to exactly one artist vertex. This pattern is referred to as a *permutation*.

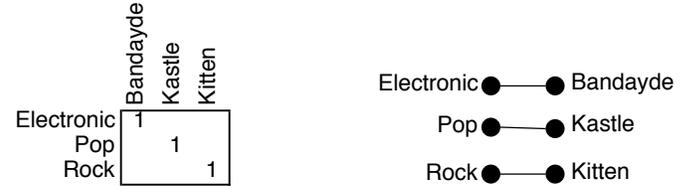

Figure 5. Special array and graph whereby each row corresponds to exactly one column. This pattern is referred to as a *permutation*.

Modifying Figure 4 by adding relationships (see Figure 5) produces a special array whereby each row has a relationship with every column. Likewise, the graph of these relationships shows the same pattern, and each genre vertex is connected to all artist vertices. This pattern is referred to as a *clique*.

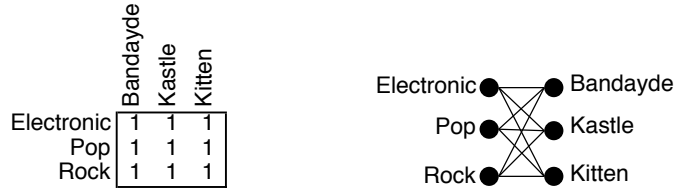

Figure 6. Special array and graph whereby each row has a relationship with every column. This pattern is referred to as a *clique*.

In addition, to the permutation and the clique pattern, there are a variety of other patterns that are important because of their special properties. Understanding how these patterns manifest themselves in associative arrays makes it possible to recognize these special patterns in spreadsheets, databases, matrices, and graphs. In a data processing system, recognizing that the data is one of these special patterns can often be used to eliminate or simplify a data processing step. For example, data with the permuation pattern shown in Figure 4 makes it very simple to look up an artist given a specific genre or a genre given a specific artist.

## VI. NULL SPACE, UNIQUENESS, AND STRETCHING

In many respects associative arrays are a generalization of matrices and inherit many of the useful behaviors that are found in matrices.

One important property of associative arrays is the circumstances under which associative array products will produce a result that contains only zeros. Recognizing these conditions can be used to eliminate steps in a data processing system. Formally this is referred to as the *null space* of the matrix.

Another important property is the conditions under which associative array products will produce a result that is not unique. If multiplying by certain classes of associative arrays always produces the same result, this property can also be used to eliminate steps in a data processing system.

Knowing when associative array products produces a zero or unchanging result is very useful for simplifying a data processing system, but these situations don't always occur. If they did, associative array products would be of little use. A situation that occurs more often is when associative array products produces a result that *stretches* one of the associative arrays by a fixed amount. It is often the case that a more complex processing step can be broken up into a series of simple stretching operations on the data, which can be used to simplify a data processing system. The directions along which a matrix will stretch are referred to as the *eigenvectors* of the matrix.

## VII. Summary

Different steps of a data processing system impose different views on the data: spreadsheets, databases, matrices, and graphs. The mathematical structure of data has many common features. Associative arrays provide a mathematically rigorous means for representing data and operations across these steps. Associative arrays can be used to swap, reorder, simplify, and eliminate steps in a data processing system.

## References

[Balazinska et al 2009] M. Balazinska, J. Becla, D. Heath, D. Maier, M. Stonebraker & S. Zdonik, *A demonstration of SciDB: A science-oriented DBMS*, Cell, 1, a2. (2009)

[Buluc et al 2014] A. Buluc, G. Ballard, J. Demmel, J. Gilbert, L. Grigori, B. Lipshitz, A. Lu- gowski, O. Schwartz, E. Solomonik & S. Toledo, *Communication-Avoiding Linear-Algebraic Primitives for Graph Analytics*, IPDPS Graph Algorithms Building Blocks (GABB), May 2014

[Chang et al 2008] F. Chang, J. Dean, S. Ghemawat, W. Hsieh, D. Wallach, M. Burrows, T.Chandra, A. Fikes & R. Gruber, *Bigtable: A Distributed Storage System for Structured Data*, ACM Transactions on Computer Systems, Volume 26 Issue 2, June 2008

[Codd 1970] E. Codd, *A Relational Model of Data for Large Shared Data Banks*, Communications of the ACM (Association for Computing Machinery) 13 (6): 37787, June, 1970

[DeCandia et al 2007] G. DeCandia, D. Hastorun, M. Jampani, G. Kakulapati, A. Lakshman, Alex Pilchin, S. Sivasubramanian, P. Vosshall & W Vogels, *Dynamo: amazons highly available key-value store*, Symposium on Operation Systems Principals (SOSP), 2007

[George 2011] L. George, *HBase: The Definitive Guide*, O'Reilly, Sebastopol, California, US,2 011

[Gilbert, Moler & Schreiber 1992] J. Gilbert, C. Moler & R. Schreiber, *Sparse matrices in MATLAB: design and implementation*, SIAM Journal on Matrix Analysis and Applications 13.1 (1992): 333-356

[Gilbert 2014] J. Gilbert, *Examples and Applications of Graph Algorithms in the Language of Linear Algebra*, IPDPS Graph Algorithms Building Blocks (GABB), May 2014

[Kallman et al 2008] R. Kallman, H. Kimura, J. Natkins, A. Pavlo, A. Rasin, S. Zdonik, E. Jones, S. Madden, M. Stonebraker, Y. Zhang, J. Hugg & D. Abadi, *H-store: a high-performance, distributed main memory transaction processing system*, Proceedings of the VLDB Endowment VLDB Endowment, Volume 1 Issue 2, August 2008 Pages 1496-1499

[Kepner 2011p] J. Kepner, *Multidimensional Associative Array Database* U.S. Patent 8,631,031 B1, filed Jan 19, 2001, awarded Jan 14, 2014

[Kepner 2012] J. Kepner, *Spreadsheets, Big Tables, and the Algebra of Associatve Arrays*, MAA & AMS Joint Mathematics Meeting, Jan 4-7, 2012, Boston, MA

[Kepner et al 2012] J. Kepner, W. Arcand, W. Bergeron, N. Bliss, R. Bond, C. Byun, G. Condon, K. Gregson, M. Hubbell, J. Kurz, A. McCabe, P. Michaleas, A. Prout, A. Reuther, A. Rosa & C. Yee, *Dynamic Distributed Dimensional Data Model (D4M) Database And Computation System*, ICASSP Special Session on Signal & Information Processing for "Big Data"; March 25-30, 2012

[Kepner & Chaidez 2013] J. Kepner & J. Chaidez, *The Abstract Algebra of Big Data*, Union College Mathematics Conference, Oct 2013, Schenectady, NY

[Kepner & Chaidez 2014] J. Kepner & J. Chaidez, *The Abstract Algebra of Big Data and Associative Arrays*, SIAM Meeting on Discrete Math, Jun 2014, Minneapolis, MN

[Kepner & Gadepally 2014] J. Kepner & V. Gadepally, Adjacency Matrices, Incidence Matrices, Database Schemas, and Associative Arrays, IPDPS Graph Algorithms Building Blocks (GABB), May 2014

[Kepner & Gilbert 2011] J. Kepner & J. Gilbert (editors), Graph Algorithms in the Language of Linear Algebra, SIAM Press, Philadelphia, 2011

[Kepner & Jansen 2015] J. Kepner & H. Jansen, *Mathematics of Big Data: Spreadsheets, Databases, Matrices, and Graphs*, SIAM Press, 2015, submitted

[Lakshman & Malik 2010] A. Lakshman & P. Malik, *Cassandra: a decentralized structured storage system*, ACM SIGOPS Operating Systems Review, Volume 44 Issue 2, April 2010

[Mattson et al 2013] T. Mattson, D. Bader, J. Berry, A. Buluc, J. Dongarra, C. Faloutsos, J. Feo, J. Gilbert, J. Gonzalez, B. Hendrickson, J. Kepner, C. Leiserson, A. Lumsdaine, D. Padua, S. Poole, S. Reinhardt, M. Stonebraker, S. Wallach, & A. Yoo, *Standards for Graph Algorithm Primitives*, IEEE High Performance Extreme Computing (HPEC), Sep 2013

[Mattson 2014] T. Mattson, *Motivation and Mathematical Foundations of the GraphBLAS*, IPDPS Graph Algorithms Building Blocks (GABB), May 2014

[Moler 1980] C. Moler, *Matlab users guide*, Alberquerque, USA (1980)

[Moler 2008] C.Moler, *Numerical computing with MATLAB*, SIAM, Philadelphia, 2008

[Smith et al 2009] M. Smith, B. Shneiderman, N. Milic-Frayling, E. Rorigues, V. Barash, C. Dunne, T. Capone, A. Perer & E. Gleave, *Analyzing (social media) networks with NodeXL*, Proceedings of the Fourth International Conference on Communities and Technologies (ACM): 255264 (2009)

[Stonebraker et al 1976] M. Stonebraker, G. Held, E. Wong & P. Kreps, *The design and implementation of INGRES*, ACM Transactions on Database Systems (TODS), Volume 1 Issue 3, Sep 1976, Pages 189-222

[Stonebraker et al 2005] M. Stonebraker, D. Abadi, A. Batkin, X. Chen, M. Cherniack, M. Fer- reira, E. Lau, A. Lin, S. Madden, E. O'Neil, P. O'Neil, A. Rasin, N.Tran & S. Zdonik, *C-store: a column-oriented DBMS*, VLDB '05 Proceedings of the 31st international conference on Very large data bases, Pages 553 - 564

[Stonebraker & Weisberg 2013] M. Stonebraker & A. Weisberg, *The Volt DB Main Memory DBMS*, IEEE Data Eng. Bull. 36.2 (2013): 21-27

[Wall, Cordova & Rinaldi 2013] M. Wall, A. Cordova & B. Rinaldi, *Accumulo Application Development, Table Design, and Best Practices*, O'Reilly, Sebastopol, California, US, 2013

[Wu et al 2014] S. Wu, V. Gadepally, A. Whitaker, J. Kepner, B. Howe, M. Balazinska & S. Madden, *MIMICViz: Enabling Visualization of Medical Big Data*, Intel Science & Technology Cen-ter retreat, Portland, OR, August, 2014